\documentclass{aip-cp}

\usepackage[numbers]{natbib}
\usepackage{rotating}
\usepackage{graphicx}

\begin{document}

\title{Observation of \boldmath{$\gamma p \to \Lambda \bar{\Lambda} p$} with GlueX at Jefferson Lab}

\author[aff2,aff1]{Reinhard Schumacher\corref{cor1}} 
\affil[aff2]{For the GlueX Collaboration}
\affil[aff1]{Department of Physics, Carnegie Mellon University, Pittsburgh, PA, 15213, USA.}
\corresp[cor1]{Corresponding author: schumacher@cmu.edu}
\eaddress[url]{https://www.cmu.edu/physics/people/faculty/schumacher.html}
\author[aff1]{Hao Li}


\maketitle

\begin{abstract}
For the first time, baryon-antibaryon photoproduction in the reaction  $\gamma p \to \Lambda \bar{\Lambda} p$  has been observed at photon energies from threshold near 4.9 GeV to 11.6 GeV.  The measurements are in progress with the GlueX spectrometer in Hall D at Jefferson Lab.   We describe here the apparatus and methods used to make these measurements and outline the physics goals of the work.   Some of the newly-seen reaction phenomenology is presented.

\noindent
(From an invited talk to the ``13th International Conference on Hypernuclear and Strange Particle Physics (HYP2018),  6-26-18'')
\end{abstract}

\section{INTRODUCTION}
The GlueX experiment in Hall D at the Thomas Jefferson National Accelerator Facility was designed to further understand the nature of confinement in Quantum Chromodynamics (QCD).   The main avenue toward this goal is a careful mapping of the light-quark meson spectrum.   It is believed that the soft gluon field in QCD can give rise to ``hybrid" mesons in which the glue is excited to carry part of the quantum numbers ($J^{PC}$) of the states~\cite{Meyer:2015eta,Dudek:2013yja}.   This would be revealed by either an overpopulation of particular spin-parity multiplets, or by the observation of quantum states excluded in the standard quark picture based on flavor-spin-rotation symmetry and the standard model gauge symmetries of $SU(3)\times SU(2)\times U(1)$.   

The experiment exploits photoproduction from a fixed hydrogen target, using both linearly polarized and unpolarized photons up to about 11.6 GeV.  The apparatus is optimized to detect charged and neutral pions and kaons, photons, and recoiling protons.   However, we have found that the design of GlueX is also favorable for the detection of all-baryonic final states.    In particular, a recent investigation has shown that two baryon-antibaryon final states can be detected with reasonable acceptances, and we are studying the reactions
\begin{equation}
\gamma + p \to p \bar{p}+ p
\label{eq:pbarp} 
\end{equation}
and
\begin{equation}
\gamma + p \to \Lambda \bar{\Lambda} + p.
\label{eq:lambarlam} 
\end{equation}
This pair of reactions is interesting to compare since their production mechanisms may be assumed to be similar but not identical.   Angular correlations of the produced pairs may offer clues to their production process, and momentum or invariant mass distributions of the produced pairs in their pair rest frames may lead to insight about their interaction potentials.    For instance, looking for  ``baryonium'' in the $p\bar{p}$ system has a long history at CERN and elsewhere~\cite{Klempt:2002ap}, albeit with no continuum quasi-bound states found.   We will have the possibility of looking for structure in the $\Lambda \bar{\Lambda}$ and $p \bar{\Lambda}$ systems.   The self-analyzing weak decay of the hyperons will allow investigations of the spin correlations in the produced pairs, something that is not practicable in the proton-anti-proton case.    

\begin{figure}[htpb]
  \centerline{\includegraphics[width=200pt]{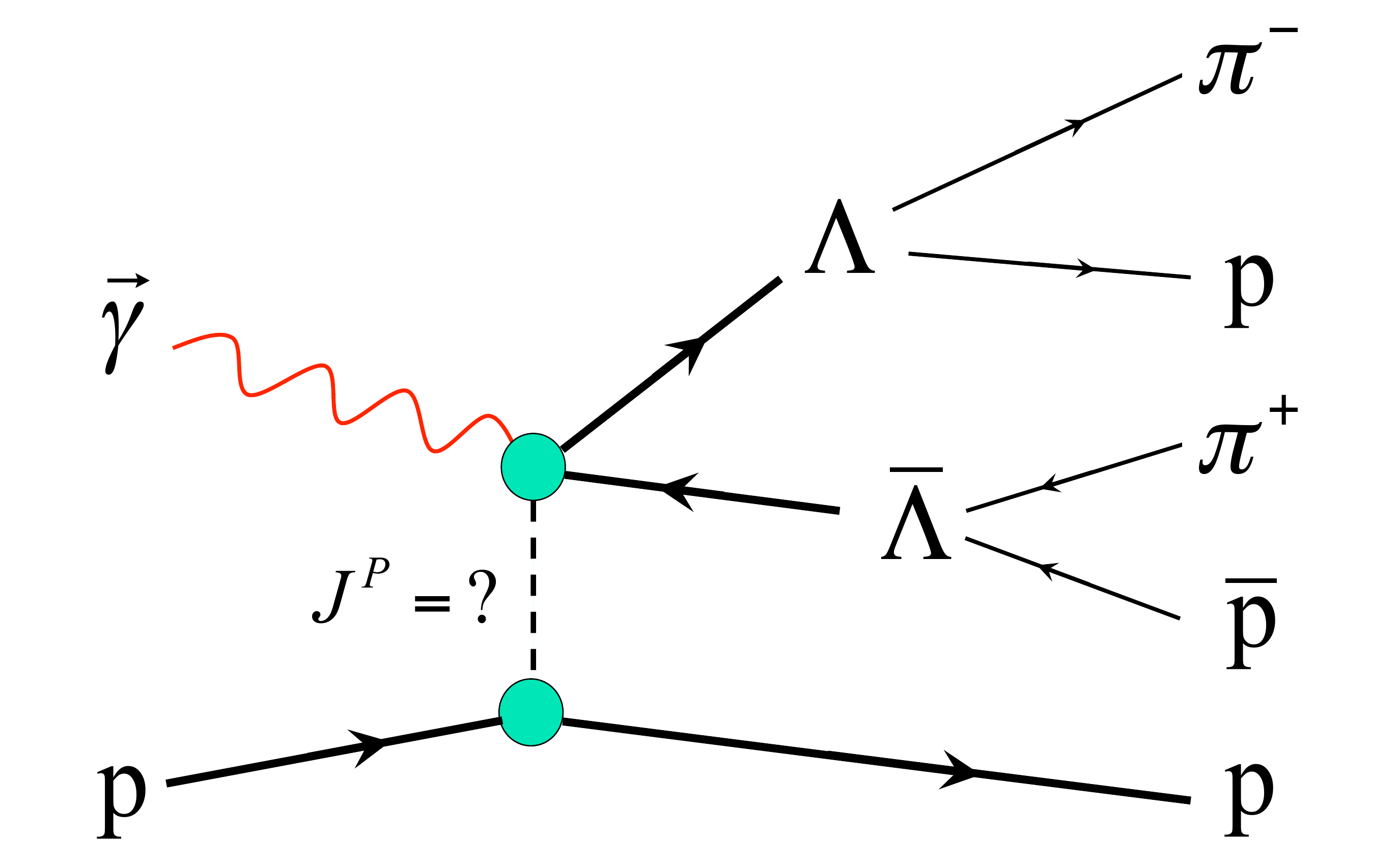}}
  \caption{One possible route for photoproduction of Lambda-anti-Lambda pairs as observed in GlueX.   The actual reaction mechanism is presently unknown. }
  \label{fig:cartoon}
\end{figure}

There seems to be very little theoretical work done on photoproduction of either $\bar{p}p$ or $\bar{\Lambda}\Lambda$ pairs.  At GlueX energies one expects that most reactions proceed through a peripheral process dominated by $t$-channel exchange of mesons, Pomerons, or some sort of Reggeons.   This is illustrated in Fig.~\ref{fig:cartoon} for the case of hyperon pair production.   We refer the reader to \citet{Gutsche:2017xtm} for the only model calculations available at this time.   It allows both vector $(\rho, \omega)$ and axial vector $(b_1, h_1)$ mesons as well as Reggeons in the $t$-channel.   It further assumes that the baryon pair is the decay product of intermediate scalar mesons $f_0(1370)$, $f_0(1500)$ and $f_0(1710)$  at the upper vertex shown in Fig.~\ref{fig:cartoon}.   The two reactions are treated in the same way, apart from the couplings to the final state baryons, leading to very similar  predictions for the $t$ dependence of the cross section and the beam spin asymmetry, $\Sigma$.  The posited intermediate scalars are all sub-threshold for the two reactions, which suggests that there could be, or ought to be, more intermediate states at higher masses.    Experiment will tell.

Thus, at the present time it is not known what the production mechanism is, including whether the hyperon pair is produced in the decay of intermediate mesonic state(s) at the upper vertex.   If they were produced from the decay of intermediate state(s), then indeed it would be of interest to look at all available  spin observables.    This includes the beam spin asymmetry, $\Sigma$, that is accessible with linearly polarized photons.  This is determined using the azimuthal event asymmetry around the plane of photon polarization.   The first publication by GlueX was  reported by  ~\citet{AlGhoul:2017nbp}, a measurement of the $\Sigma$ observable for $\pi^0$ and $\eta$ production with about 100 times the statistics of the only previous measurements at SLAC, from decades ago.   This observable can also be measured for the hyperon reaction;  we are not presenting preliminary results for this observable here, however, since event statistics are not yet sufficient.

Furthermore, the self-analyzing weak decay of the hyperons gives access to the spin polarization $P_\Lambda$, of the created particles.   There are no model predictions for this, but certain statements can be tested.  $C$-parity demands that we have in pair production $P_\Lambda = P_{\bar{\Lambda}}$.   Strong $CP$ invariance demands that the weak decay asymmetries of the pair are opposite:  $\alpha = - \bar{\alpha}$.  CPT invariance requires the lifetimes of the pair to be equal: $\tau_\Lambda =  \tau_{\bar{\Lambda}}$.  The experimental (PDG) limit on the latter quantity is only about 1\%, and GlueX may be able to improve upon it if sufficient statistics are accumulated.

Not only may the individual $\Lambda$ and $\bar{\Lambda}$ particles be polarized, subject to the requirement of C-parity, but even if they are not, there may be a measurable correlation between the two spins when the particles are formed together.   For example, such correlations have been studied extensively for the reaction $p \bar{p} \to \Lambda \bar{\Lambda}$ at LEAR~\cite{Klempt:2002ap,Barnes:1996si}, and the formalism of spin correlations is well developed~\cite{PhysRev.135.B540,Tabakin:1985yv,Lyuboshitz:2010zz} .   The basic idea is to test the extent to which the pair is produced in a spin triplet state versus a spin singlet state.   In the spin-1/2 Pauli representation of hyperon spin, the full correlation matrix of producing pair $\{\bar{m},n\}$, where $m$ and $n$ index the three coordinate directions chosen in the pair rest frame,  is written 
\begin{equation}
C_{\bar{m}n} = \left<\vec{\sigma}_{\bar{m}} \times \vec{\sigma_n}\right >,
\label{eq:correlation} 
\end{equation}
where the bracket denotes the expectation value.   This can be evaluated experimentally as a suitable correlated sum over the decay proton -  antiproton distributions.   As shown in the given references, $C$-parity and spatial parity $P$  result in several components of this matrix being zero.   The so-called singlet fraction, $S_F$, the fraction of events in which the hyperon pair in produced in a spin singlet state, is defined by
\begin{equation}
S_F =\frac{1}{4} \left  ( 1- \left<\vec{\sigma}_\Lambda \cdot  \vec{\sigma}_{\bar{\Lambda}}\right> \right ) = \frac{1}{4} \left ( 1 + C_{\bar{x}x} - C_{\bar{y}y} + C_{\bar{z}z}    \right)
\label{eq:singletfraction} 
\end{equation}
Measurement of these quantities are a hope for the future at GlueX when enough statistics have been accumulated.  However, we would like to encourage model-based estimates of what might be expected for these observables.

The reaction in Eq.~\ref{eq:pbarp}  has been investigated by the CLAS Collaboration up to about 5 GeV photon energy~\cite{Phelps}, and is being  explored at GlueX energies, but it is not the topic of this article.   We will discuss here the preliminary information we have obtained for the reaction in Eq.~\ref{eq:lambarlam},   {\it i.e.} the hyperon case.

\section{APPARATUS AND PROCEDURES}
The GlueX experiment has been taking analysis-quality physics data since the spring of 2016.   The preliminary results shown in this contribution stem from data obtained  during the 2016 and 2017 running periods.   

\subsection{The GlueX Detector}
\begin{figure}[htpb]
  \centerline{\includegraphics[width=400pt]{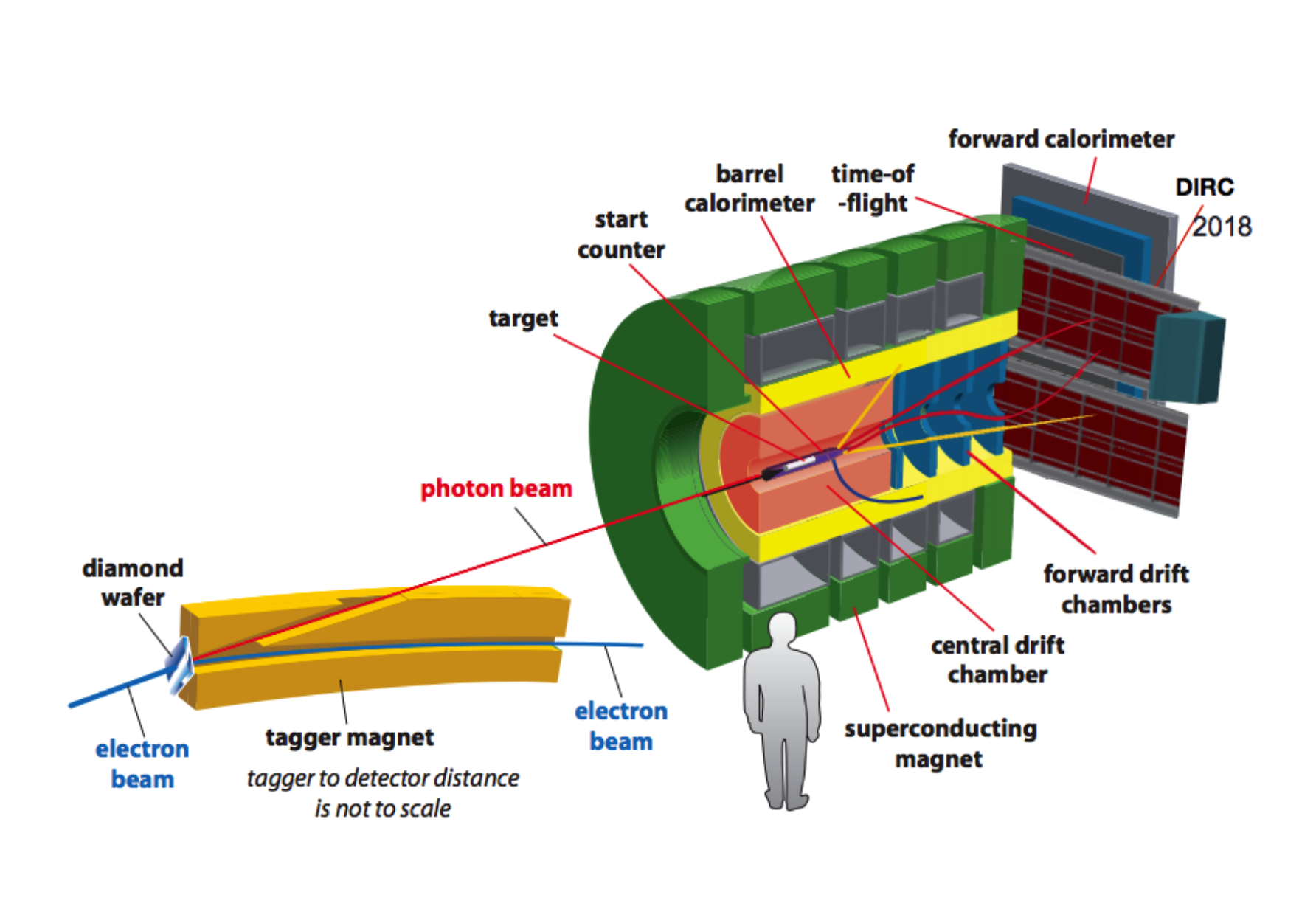}}
  \caption{Schematic of the GlueX detector and the photon beam tagger. }
  \label{fig:gluex}
\end{figure}

The GlueX detector in Hall D at the CEBAF accelerator at Jefferson Lab is illustrated in Fig.~\ref{fig:gluex}.  For these measurements,  an 11.6 GeV electron beam impinges on a diamond radiator which produces a ``coherent''  Bremsstrahlung real photon beam with a substantial linear polarization in a narrow energy range peaking at 9 GeV.   After collimation, the degree of polarization in the coherent peak is about 40\%, while the rest of the energy spectrum is unpolarized.  Recoiling electrons are analyzed using a dipole ``tagging" magnet and a fine-grained hodoscope scintillator array to define the photon energy.   The beam quality,  flux, and polarization are monitored  upstream of the main detector using an $e^+e^-$ pair spectrometer and a triplet polarimeter (not shown).   A  ${\sim}30$~cm liquid hydrogen physics target is located in the 2 Tesla solenoidal magnetic field.    Charged particles produced in this fixed-target setup are mostly forward-directed in the laboratory frame.   Charged particles are tracked by a straw-tube central drift chamber with axial geometry but including small-angle ($6^\circ$) ``stereo" layers.   Planar drift chambers in the forward direction further track charged particles.   The time of flight (TOF) of forward-going charged massive particles is measured using a scintillator Start Counter that surrounds the target combined with a forward TOF scintillator wall.   Photons  and electrons are detected in a lead/scintillator fiber Barrel Calorimeter inside the magnet and a lead-glass Forward Colorimeter.   These latter two detectors also detect the arrival of charged particles;  the trigger for the event readout requires a minimum energy to be deposited in the calorimeters.    

The angular coverage of the detector is complete down to opening angles from the beam of  about two degrees.   Most reactions are strongly forward-peaked in the lab frame. Tracking and photon detection extend back to about $128^\circ$.  Pions are reconstructed for momenta down to 100 MeV/c and protons to 300 MeV/c.   The momentum resolution per track range from 1\% to 3\%.  Thus, we say the detector is close to hermetic in kinematic coverage for our purpose.

\subsection{Data Analysis}
The fully exclusive final state for this reaction consists of five changed particles
\begin{equation}
\gamma + p \to \Lambda + \bar{\Lambda} + p \to \{\pi^- +  p\}  + \{\pi^+  + \bar{p}\} + p,
\end{equation}
as illustrated in Fig.~\ref{fig:cartoon}.    Events with at least five charged tracks were selected.  Kinematic fitting was used on all possible particle identification combinations of these tracks and using all in-time photons in the tagger.  Events for which the fits simply converged were kept as candidates.  Two fits were done for each combination.  The first constrained overall energy and momentum conservation and had constraints on the primary and secondary (decay) vertices.  The second fit likewise  constrained overall energy and momentum in the event, the three vertices, but in addition constrained the hyperon masses.   After the kinematic fitting, further particle identification (PID) cuts were made on individual tracks to separate pions from protons.  These included {\it dE/dx} cuts in the central drift chamber, where pions and protons could be separated  up to about 1 GeV/c.  Timing cuts between the target and TOF wall were made, where pions and protons could be separated up to about 5 GeV/c.   The four nanosecond photon arrival-time microstructure of the CEBAF beam in Hall D was used to help select the best beam photon candidate for each event.

\begin{figure}[htpb]
  \centerline{\includegraphics[width=400pt]{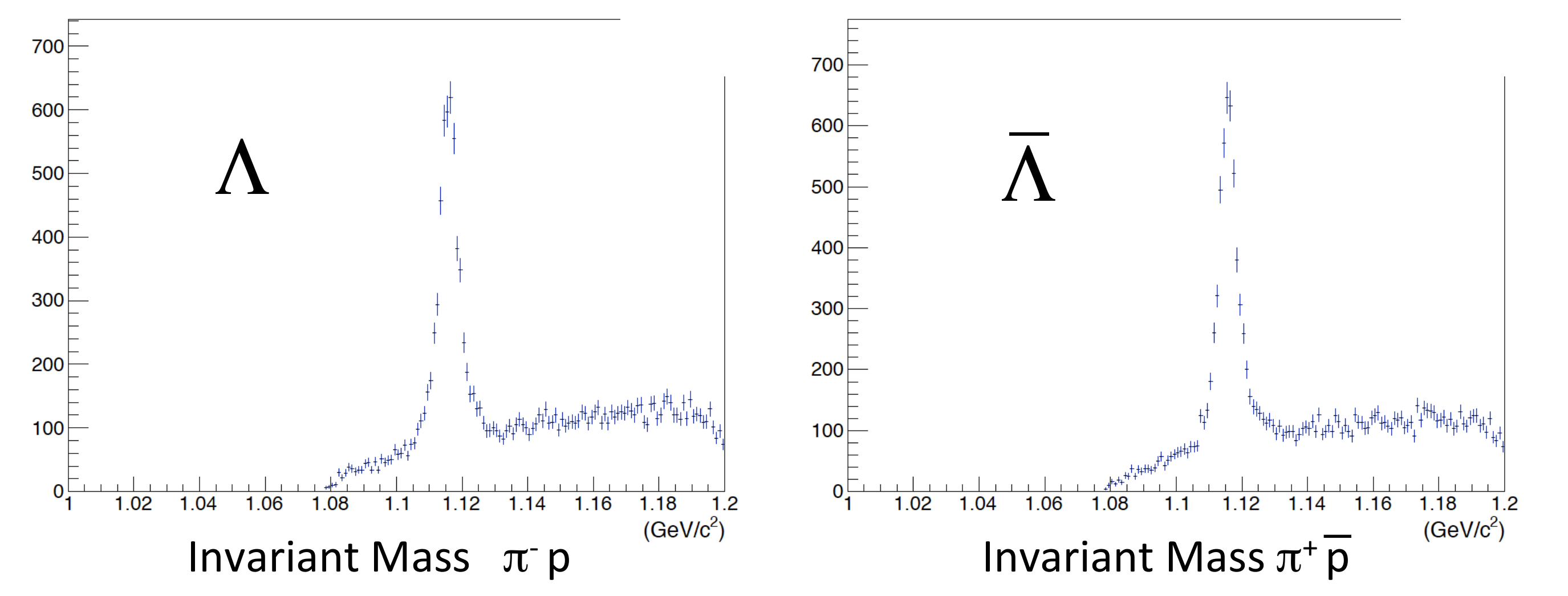}}
  \vspace*{+10cm}
  \caption{The preliminary invariant mass of the $\Lambda$ from the detected $\pi^-$ and $p$, and the corresponding distribution for the $\bar{\Lambda}$ from $\pi^+$ and $\bar{p}$.  The kinematic fit used to produce these distributions did not constrain the hyperon masses.  The figure shows that the hyperons produce clear signals in the GlueX detector.}
  \label{fig:invariantmass}
\end{figure}

The distributions of invariant masses for the reconstructed $\Lambda \to \pi^- p$ and the $\bar{\Lambda} \to \pi^+ \bar{p}$ states are shown in Fig.~\ref{fig:invariantmass}.   At this stage, all {\it dE/dx} and timing cuts were applied to individual tracks and a 1\%  confidence level cut was made on the kinematic fit result.    The hyperon masses were not constrained in this fit, so one sees that the hyperons can be clearly distinguished in GlueX, and that the signals look very similar to each other, as would be expected.    Separate ``$\pm 3$ sigma'' cuts on each state were made, or 18 MeV in total, but the remaining background under the states due to beam-tagging and hadronic particle mis-identification has not been subtracted in the subsequent plots.  The number of signal events in this preliminary analysis is only about 3700;  as a consequence of this, these spectra are summed over all photon beam energies (see next section).  

\section{PRELIMINARY RESULTS}

\subsection{Angular Distribution}
\begin{figure}[htpb]
  \centerline{\includegraphics[width=400pt]{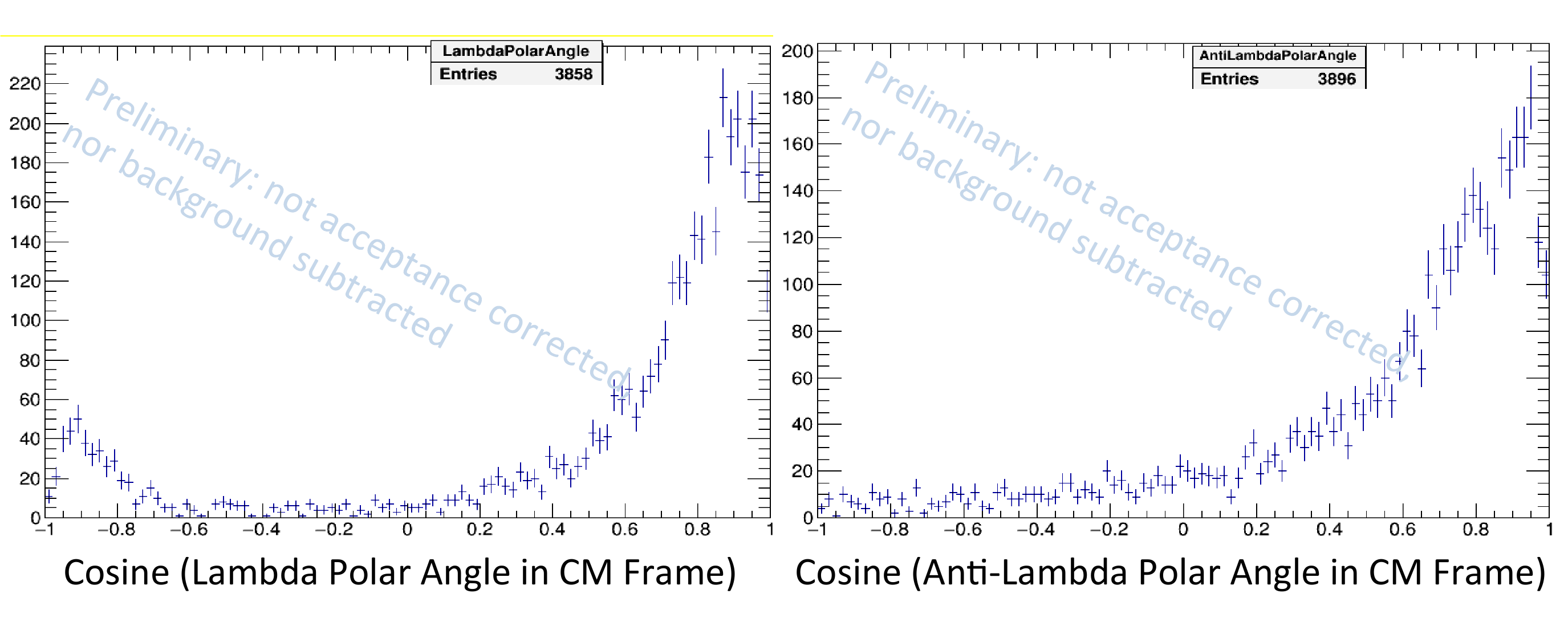}}
  \vspace*{-10cm}
  \caption{The overall center-of-mass frame angular distribution of events of the $\Lambda$ particles (left) and the corresponding distribution for the $\bar{\Lambda}$ particles (right).  One notes that both hyperons are forward peaked, but that the $\Lambda$ particles also have a peak in the backward direction.   These spectra have not been acceptance corrected or background subtracted.}
  \label{fig:costhetalambdacm}
\end{figure}

The first result concerns the distribution of hyperons in the overall beam-target ($\gamma p$) center-of-mass (CM) frame for the reaction.   These are shown in Fig.~\ref{fig:costhetalambdacm}.  The two distributions would look the same if the diagram in Fig.~\ref{fig:cartoon} were the whole story:  the hyperons would be produced  ``symmetrically'' in their pair rest frame, due to parity conservation, and projected into the CM frame with the same angular distribution.    Instead, both distributions have a forward peak, but the $\Lambda$ distribution has a second peak at back angles.   A caveat here is that these spectra have not yet had beam-tagging accidentals removed (which is thought not to affect the spectral shapes), nor has the ``sideband'' background seen in Fig.~\ref{fig:invariantmass} been carefully removed (as mentioned above).   However, we have checked (not shown here) that the background under the hyperon signal is not backward peaked.   

\begin{figure}[htpb]
  \centerline{\includegraphics[width=400pt]{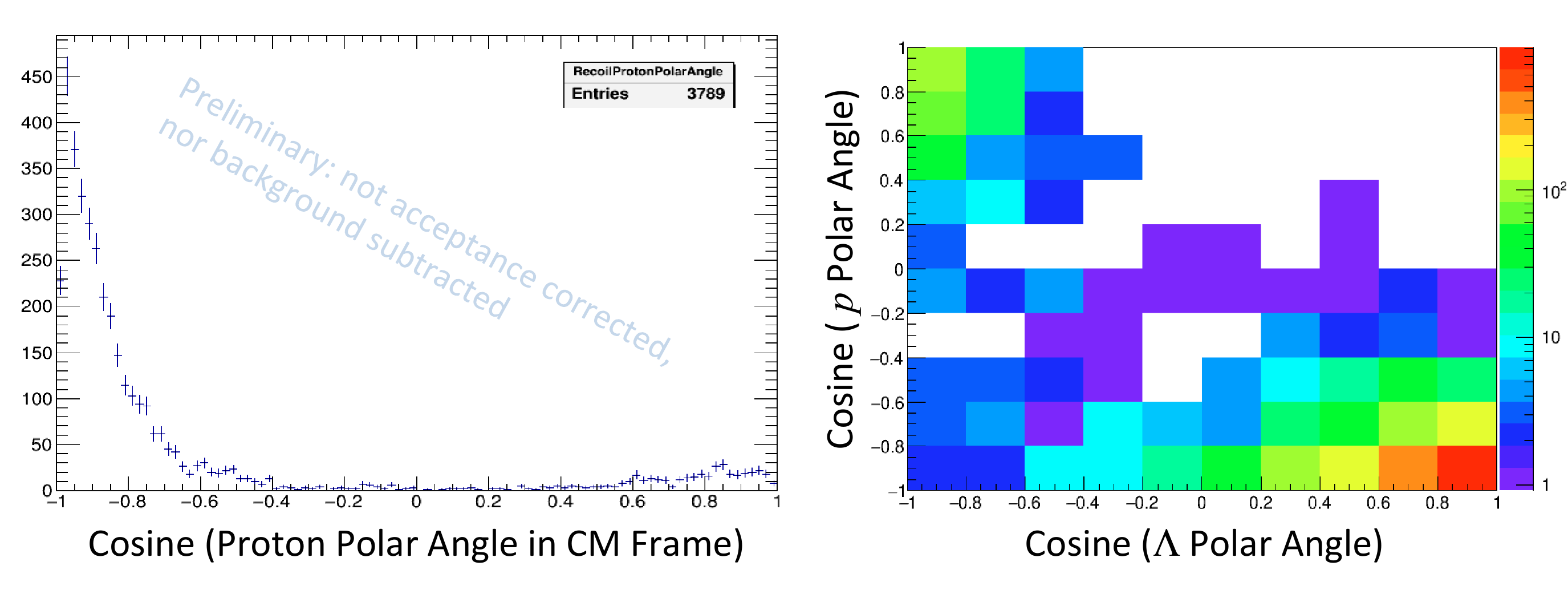}}
  \vspace*{-10cm}
  \caption{
  Left: The overall center-of-mass frame angular distribution of events of the final state proton.  The protons are backward peaked, corresponding to forward-going $\Lambda$ particles, while the second proton peak in the forward direction goes with backward-going $\Lambda$  particles.     Right: The correlation of the protons in the events with the $\Lambda$ particles, showing two clusters of events.
   }
  \label{fig:costhetaprotoncm}
\end{figure}

This difference hints that at least two reaction mechanisms are in play in this reaction.   Further evidence of this is provided by the CM angular distribution of the remaining final-state proton, which is shown in Fig.~\ref{fig:costhetaprotoncm}.  These protons are backward peaked  with a second peak in the forward direction, mirroring the $\Lambda$ distribution in Fig.~\ref{fig:costhetalambdacm}.   Indeed, the backward protons are correlated with the forward $\Lambda$ particles.

\begin{figure}[htpb]
  \vspace{-15.cm}
  \centerline{\includegraphics[width=300pt]{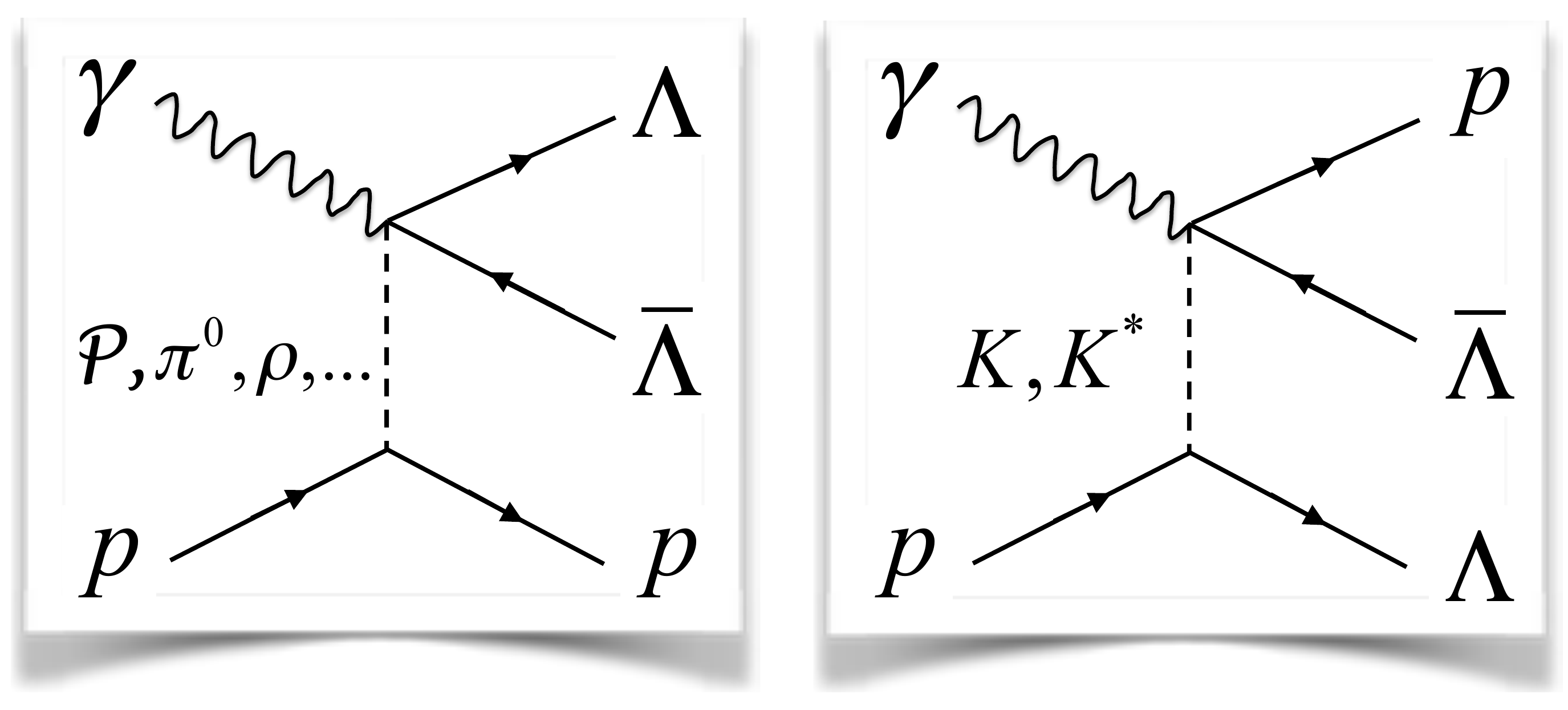}}
  \vspace{-15.cm}
  \caption{Alternative production mechanisms that involve non-strange  (left) or strangeness (right) exchange in the $t$ channel.}
  \label{fig:feynman}
\end{figure}

A natural way to explain this situation is to envision the two Feynman diagrams shown in Fig.~\ref{fig:feynman}.   In the first instance, a non-strange exchange occurs in the $t$ channel, possibly including the Pomeron, $\pi$, and $\rho$.  This creates the hyperon and anti-hyperon together at the upper vertex which, as a pair, would be dominantly seen at low $t$ and be symmetrically distributed.   In the second instance, a strange exchange occurs in the $t$ channel, which converts the target proton into a recoiling $\Lambda$ that appears at back angles in the CM frame.   Meanwhile, at the top vertex a $p \bar{\Lambda}$ pair is created at low $t$ that appears at small CM angles.   We have checked that the $t$ dependence of the respective pairs looks exponential, {\it i.e.} goes as $e^{bt}$, but before proper background subtraction and acceptance corrections we do not show these distributions here.    It would be very interesting to have model predictions to test this phenomenology.   

\begin{figure}[htpb]
  \centerline{\includegraphics[width=400pt]{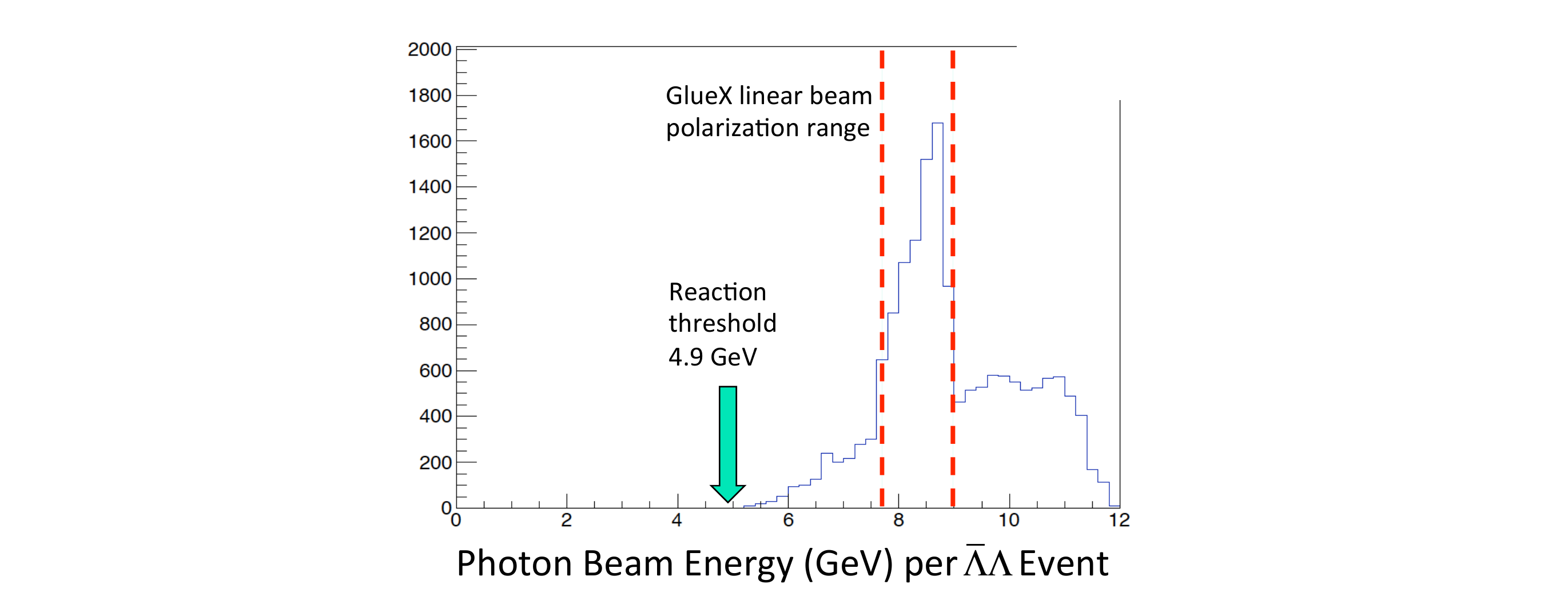}}
  \vspace*{-10cm}
  \caption{
The distribution of photon energies for reconstructed hyperon events.   The peak of events between 8 GeV and 9 GeV  is the increased photon intensity recorded in the coherent Bremsstrahlung peak where the photons are linearly polarized.  }
  \label{fig:beamenergy}
\end{figure}

Figure~\ref{fig:beamenergy} shows the photon energy distribution for all events in the final $\Lambda \bar{\Lambda}$ spectra.   The big peak near 9 GeV shows the range of polarized photons in GlueX and is not indicative of any physics process.   The full photon intensity in the beam outside the peak is much flatter than shown, which indicates qualitatively that the acceptance and (probably) the cross section starts small near threshold and grows with photon energy.  In the range from 8~GeV to 9~GeV the beam spin asymmetry, $\Sigma$, can be measured.

\subsection{Invariant Mass of Pairs}

\begin{figure}[htpb]
  \centerline{\includegraphics[width=400pt]{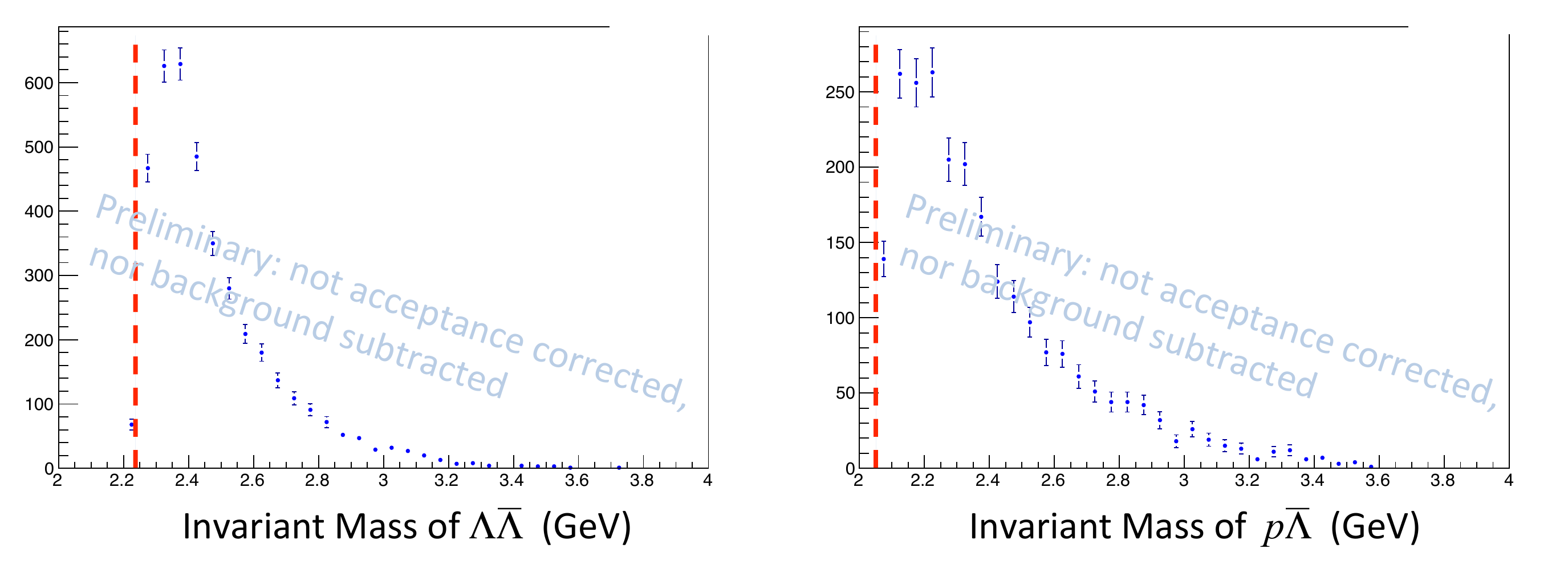}}
  \vspace*{-10cm}
  \caption{
  (Left) The invariant mass of the produced ($\Lambda\bar{\Lambda}$) pair when the $\Lambda$ particle goes forward in the CM with the $\bar{\Lambda}$.
  (Right) The invariant mass of the produced proton and the produced  $\bar{\Lambda}$ when the $\Lambda$ particle goes backward in the CM frame ($p\bar{\Lambda}$).}
  \label{fig:onium}
\end{figure}

Since the events fall into two categories, ${\Lambda\bar{\Lambda}}$ and ${p \bar{\Lambda}}$ pairs, it is interesting see what the invariant mass distributions of these pairs looks like;  this is shown in Fig.~\ref{fig:onium}.  The analogous spectrum for $p\bar{p}$  production was scrutinized for many years for signs of ``protonium"  at CERN and elsewhere~\cite{Klempt:2002ap} with no positive results.  Both distributions rise very sharply from their respective thresholds and form a smooth distribution.   To what extent these distributions are controlled by the small-$t$ nature of the production mechanisms is under investigation.   As a next step, we will look into various kinematic models of the production process using Monte Carlo simulations to glean more insight into these spectra.   In any case, with present statistics one does not observe any very narrow structures that would be candidates of an ``-onium" system in the $\Lambda \bar{\Lambda}$ and $p \bar{\Lambda}$ final states. 

One can look, equivalently, at the momentum distribution in the pair rest frames normalized to phase space.   In looking for near-threshold interactions, this is another way to investigate these baryon pairs that may prove fruitful~\cite{fabbietti}.

\subsection{Further Discussion and Other Strangeness Physics at GlueX}

When the GlueX kinematic fitter constrained the mass of the hyperons, it also produced 3-space vertices for these particles in addition to a primary photoproduction vertex.   From these and from the momentum of the particles, one computes the lifetime distribution of the hyperon and the anti-hyperon in their respective rest frames.   The goal of this measurement is to compare the lifetimes, which, as mentioned earlier, should be equal according to CPT invariance.   The PDG limit for this equality for the $\Lambda$ is a rather modest 1\%, which GlueX can, with sufficient data, improve upon.    

The GlueX program has other measurements under analysis that are related to strangeness physics.   These include
\begin{itemize}
\item Measuring the beam spin asymmetry for the elementary $\gamma p \to K^+ \Lambda$ reaction;
\item Investigating photoproduction of the $\Lambda(1520)$ and higher-mass hyperons;
\item Looking in photoproduction for $\Xi$ states beyond the $\Xi(1320)$ and the $\Xi(1530)$.
\end{itemize}
Also, to extend the momentum range of $K/\pi$ separation in the system, a differential Cherenkov counter DIRC (Detection of Internally Reflected Cherenkov light) system~\cite{Stevens:2016cia} is being installed in 2018.   It will extend the separation range from about 2.0 GeV/c (provided by the TOF) to over 4.0 GeV/c using fused silica radiators that were first used successfully by BaBar at SLAC.   This range-extension is considered crucial to fully exploit the meson-production program at GlueX.

\section{SUMMARY AND CONCLUSIONS}
The main point of this article has been to show that the photoproduction off the proton of hyperon pairs ($\Lambda \bar{\Lambda})$ has been seen for the first time at GlueX.   It has been shown that the hyperon angular distributions indicate that more than one distinct reaction mechanism is at work, and that we believe the evidence points to non-strange and strange meson exchange in the $t$ channel.   However, presently no substantive theoretical effort has been made to support this inference.   All spectra shown here are very preliminary, having yet to undergo careful background subtraction and acceptance correction.   Using the capabilities of the Hall D polarized beam and the GlueX detector, we expect to produce differential cross sections and various spin observables.   The hyperons' self-analyzing weak decays  will allow various tests of the production mechanism.    Comparison to the $p\bar{p}$ reaction will be made over the same ranges of energy and angle:  the event statistics are several orders of magnitude greater, but no spin information other than $\Sigma$ can be obtained.   We hope to have stimulated some interest from the theory community to examine this reaction in detail.  

\section{ACKNOWLEDGMENTS}
We thank Dr. Naomi Jarvis for assistance in setting up the analysis computing environment.  The initial programming help in this project of Carnegie Mellon undergraduate students Mr. Samuel Dai and Mr. Viren Bajaj is gratefully acknowledged.  The work of the Medium Energy Physics group at Carnegie
Mellon University was supported by DOE Grant No. DE-FG02-87ER40315.  The Thomas Jefferson National Accelerator Facility is supported by the U.S. Department of Energy, Office of Science, Office of Nuclear Physics under contract DE-AC05-06OR23177.


\nocite{*}
\bibliographystyle{aipnum-cp}%
\bibliography{HYP2018_Schumacher}%

\end{document}